\input{psfig.sty}
\documentclass[preprint]{aastex}

\begin{document} 
  
\title{Partially Absorbed Comptonization Spectrum from the Nearly Edge-on 
Source X~1822--371} 
 
\author{R. Iaria\altaffilmark{1}, 
T. Di Salvo\altaffilmark{2,1}, 
L. Burderi\altaffilmark{3}, 
N. R. Robba\altaffilmark{1}} 
\altaffiltext{1}{Dipartimento di Scienze Fisiche ed Astronomiche,  
Universit\`a di Palermo, via Archirafi n.36, 90123 Palermo, Italy} 
\authoremail{iaria@gifco.fisica.unipa.it} 
\altaffiltext{2}{Astronomical Institute ``Anton Pannekoek," University of 
Amsterdam and Center for High-Energy Astrophysics,
Kruislaan 403, NL 1098 SJ Amsterdam, the Netherlands}
\altaffiltext{3}{Osservatorio Astronomico di Roma, Via Frascati 33,  
00040 Monteporzio Catone (Roma), Italy} 
   
\begin{abstract} 
 
We report the results of a spectral analysis over the range 0.1--200
keV performed on the dipping source X~1822--371 observed by BeppoSAX.
We find the best fit to the continuum using a partially covered
Comptonization model, due to scattering off soft seed photons by
electrons at a temperature of $\sim 4.8$ keV, without the presence of
any soft blackbody emission.  The equivalent hydrogen column obtained
for the absorbed component is $\sim 4.5 \times 10^{22}$ cm$^{-2}$, an
order of magnitude larger than the Galactic absorption for this
source, and the covering fraction is $\sim 71\%$.  Because the
inclination angle of X~1822--371 to the line of sight is $\sim
85^{\circ}$, this model gives a reasonable scenario for the source:
the Comptonized spectrum could come from an extended accretion disk
corona (ADC), probably the only region that can be directly observed
due to the high inclination.  The excess of matter producing the
partial covering could be close to the equatorial plane of the system,
above the outer disk, occulting the emission from the inner disk and
the inner part of the ADC.

An iron emission line is also present at $\sim 6.5$ keV with an
equivalent width of $\sim 150$ eV. We argue that this strong iron line
cannot be explained as reflection of the Comptonized spectrum by the
accretion disk.  It is probably produced in the ADC.  An emission line
at $\sim 1.9$ keV (with an equivalent width of $\sim 54$ eV) and an
absorption edge at $\sim 8.7 $ keV (with an optical depth of $\sim
0.1$) are also required to fit this spectrum.  These features are
probably produced by highly ionized iron (Fe XXIV) present in the
outer part of the ADC, where the plasma density is $\sim
10^{11}-10^{12}$ cm$^{-3}$ and ionized plasma is present.

\end{abstract} 
 

\section{Introduction} 
 
Low Mass X-ray Binaries (LMXB) consist of a low mass star ($M \le 1\
M_\odot$) and a neutron star (NS), generally with a weak magnetic
field ($B \le 10^{10}$ Gauss).  In these systems the X-ray source is
powered by accretion of mass overflowing the Roche lobe of the
companion star and forming an accretion disk around the NS. Different
inclinations of the line of sight with respect to the orbital plane
can explain the different characteristics of the light curve observed
in these systems.  At low inclinations eclipses and dips will not be
visible in the light curves, while they can be present at high
inclinations.  About 10 LMXBs are known to show periodic dips in their
X-ray light curves. The dip intensities, lengths and shapes change
from source to source, and, for the same source, from cycle to cycle.
Dips are probably due to a thicker region in the outer rim of the
accretion disk, formed by the impact with the disk of the gas stream
from the Roche-lobe filling companion star.  For systems seen almost
edge-on, X-ray emission is still visible due to the presence of an
extended accretion disk corona (ADC, see White \& Holt, 1982) which
can be periodically eclipsed by the companion star.
 
X~1822--371 is a LMXB seen almost edge-on, with an inclination angle
$i \sim 85^\circ$ (Hellier \& Mason, 1989).  Its light curve shows
both dips and eclipses of the X-ray source by the companion star.  The
partial nature of the eclipse indicates that the X--ray emitting
region is extended, and that the observed X--rays are scattered in an
ADC.  The 1-10 keV spectrum of X~1822--371 as observed by EXOSAT was
fitted by an absorbed blackbody plus a power-law component, with an iron
emission line at $\sim 6.7$ keV (Hellier and Mason, 1989).  The 1-30
keV spectrum observed by {\it Ginga} could not be described by the
model above, that gave a reduced $\chi^2$ of 12 (Hellier, Mason and
Williams, 1992), probably for the better statistics and the wider
energy range of {\it Ginga} with respect to EXOSAT. Other combinations
of power-law, blackbody and thermal bremsstrahlung were used to fit
these data, but none of these models gave an acceptable fit.  Heinz \&
Nowak (2001) analysed simultaneous observations with RXTE and ASCA of
X~1822--371. They showed that both the source spectrum and light curve
can be well fitted by two models, representing the case of an
optically thick and an optically thin corona, respectively.  In the
first case, no soft thermal component from the inner region
contributes to the source spectrum and the emission from the corona is
described by a cutoff power law partially absorbed by a cold
atmosphere above the disk. In the second case the model consists of a
blackbody component, emitted from the central source and scattered
into the line of sight by the optically thin corona, and a cutoff
power law emitted by the corona. Both this models could well describe
the data, and therefore it was not possible to distinguish between
them.

Recently this source was studied by Parmar et al. (2000) using data
from the BeppoSAX satellite in the energy range 0.3--40 keV.  They fit
the spectrum using a Comptonization model, with a seed-photon
temperature of $\sim 0.1$ keV, an electron temperature of $\sim 4.5$
keV, and a Comptonizing cloud optical depth of $\tau \sim 26$, plus a
strong blackbody component, at the temperature of $\sim 1.9$ keV,
which contributes more than $40\%$ of the 0.3--40 keV flux of the
source.  An emission line at $\sim 6.5$ keV and an absorption edge at
$\sim 1.3$ keV were also present.  We have analysed the same data
using the whole BeppoSAX range (0.1--200 keV).  We confirm that the
model used by Parmar et al. (2000) can well fit the X~1822--371
spectrum, but we find a better fit using a Comptonization model with
partial covering, two emission lines, at $\sim 6.5$ keV and $\sim 1.9$
keV, respectively, and an absorption edge at $\sim 8.7$ keV. In our
model there is no need of a blackbody emission or an absorption edge
at low energy.

\section{Observations and Spectral Analysis} 

The Narrow Field Instruments (NFI) on board BeppoSAX satellite (Boella
et al. 1997) observed X~1822--371 on 1997 September 9 and 10, for an
effective exposure time of $\sim 43$ ks.  The NFIs are four co-aligned
instruments which cover more than three decades of energy, from 0.1
keV up to 200 keV, with good spectral resolution in the whole range.
The Low Energy Concentrator/Spectrometer (LECS) operating in the range
0.1--10 keV and the Medium Energy Concentrator/Spectrometer (MECS,
1--11 keV) have imaging capabilities with field of view of $20'$ and
$30'$ radius, respectively. We selected data for scientific analysis
in circular regions, centered on the source, of $8'$ and $4'$ radii
for LECS and MECS, respectively. The background subtraction was
obtained using blank sky observations in which we extracted the
background spectra in regions of the field of view similar to those
used for the source. The High Pressure Gas Scintillation Proportional
Counter (HPGSPC, 7--60 keV) and the Phoswich Detection System (PDS,
13--200 keV) are non-imaging instruments, because their field of
views, of $\sim 1^\circ$ FWHM, are delimited by collimators.  In the
spectral analysis we used the standard energy ranges for the NFIs,
that are: 0.12--4 keV for the LECS, 1.8--10 keV for the MECS, 7--30
keV for the HPGSPC and 15--200 keV for the PDS.  As customary, in the
spectral fitting procedure we allowed for different normalizations in
the LECS, HPGSPC and PDS spectra relative to the MECS spectrum, and
checked {\it a posteriori} that derived values are in the standard
range for each instrument.  We rebinned the energy spectra in order to
have at least 30 counts/channel.  The LECS and MECS spectra were
furtherly rebinned in order to oversample the full width at half
maximum of the energy resolution by a factor of 5 in the whole energy
range.\footnote{see the BeppoSAX cookbook at
http://www.sdc.asi.it/software/index.html}
 
The observed average unabsorbed flux of the source in the 0.1--100 keV  
energy range is $1.55 \times 10^{-9}$ ergs cm$^{-2}$ s$^{-1}$.  
Adopting a distance of 2.5 kpc (Mason \& Cordova, 1982), it 
corresponds to an unabsorbed luminosity of $1.15 \times 10^{36}$ ergs/s. 
This is compatible with the previously reported isotropic   
luminosity of the X-ray source of $\sim 10^{36}$ ergs/s   
(Mason \& Cordova, 1982).  

In Figure 1 (upper panel) we plotted the X~1822--371 light curve in
the energy bands 1.8--10.5 keV (MECS data) versus the orbital phase
(using the orbital period reported in Parmar et al. 2000).  In the
light curve a sinusoidal variation and a partial eclipse at the
orbital phase of $\sim 0.8$ are present.  The hardness ratio (the
ratio between the counts in the two energy bands 4--8 keV and 1--4
keV, lower panel in Fig. 1) does not show large variations.  Therefore
we performed our spectral analysis on the source spectrum averaged
over the whole orbital phase.
 
We fitted the model obtained by Parmar et al.~(2000) in the energy
range 0.3--40 keV, to the BeppoSAX data in the energy range 0.12--200
keV.  The model consists of photoelectric absorption by cold matter, a
blackbody (BB), a Comptonized component (hereafter {\it Comptt},
Titarchuk 1994), an emission line at $\sim 6.5$ keV and an absorption
edge at $\sim 1.3$ keV.  We obtained a $\chi^2/d.o.f. = 245/201$.  The
values of the parameters are compatible with the values obtained in
the range 0.3--40 keV (Parmar et al., 2000) and are reported in Table
1 (Model 1).  This model, however, presents two characteristics that,
in our opinion, are hard to explain, i.e. the amount of photoelectric
absorption, $N_H$, and the luminosity of the blackbody component. The
value of $N_H$ reported in Parmar et al. (2000) is $\sim 1.2 \times
10^{20}$ cm$^{-2}$, an order of magnitude lower than the Galactic
absorption expected in the direction of this source (see \S 3).  Using
the whole BeppoSAX energy range, we find a larger value $N_H \sim 6
\times 10^{20}$ cm$^{-2}$, that is still smaller (by a factor of 1.7)
than the expected value.  Again a very low value of photoelectric
absorption is found in the RXTE and ASCA data when the optically thin
corona model (consisting of blackbody and cutoff power law) is used
(Heinz \& Nowak 2001).  Moreover, the blackbody component in this
model contributes more than 40\% of the total source luminosity, that
is quite large considering that the source is seen almost edge-on.
Both these points were already noted and discussed by Parmar et
al. (2000).

We were therefore motivated to search for another model that could
both best fit the data and have a simple physical interpretation.
Then we repeated the analysis trying several models. In particular
given that the source is seen almost edge-on, we tried a
Comptonization model ({\it Compst}, Sunyaev \& Titarchuk, 1980) with
partial covering and an emission line at $\sim 6.5$ keV. In this way,
we obtained a $\chi^2/d.o.f. = 291/204$; the corresponding values of
the parameters are reported in Table 1 (Model 2).  In Figure 2 (upper
panel) we present the BeppoSAX broad band spectrum and Model 2, and in
the same figure (middle panel) we show the residuals in units of
$\sigma$ with respect to Model~2. In the residuals an emission feature
at $\sim 2$ keV and an absorption feature at $\sim 10$ keV are
present.  These residuals are of the order of $6-8\%$ at 2 keV and
of $5\%$ at 10 keV, much higher than any systematic residuals in the MECS
and HPGSPC data with respect to the Crab spectrum.\footnote{see the BeppoSAX
cookbook at the web site http://www.asdc.asi.it/bepposax/software/index.html.}
Therefore we firstly added an emission line at $\sim 1.9$
keV, which significantly improved the fit.  We obtained a
$\chi^2/d.o.f. = 236/201$, obtaining a probability of chance
improvement of the fit (with respect to Model~2) of $\sim 3.61 \times
10^{-9}$.  Then we added an absorption edge at $\sim 8.7$ keV,
obtaining a $\chi^2/d.o.f. = 206/199$, and a probability of chance
improvement of the fit (with respect to the previous model) of $\sim
3.51 \times 10^{-5}$.  
These two features fall at the ends of the MECS energy range. 
We note that no problems are known to exist in the MECS response matrix
at the ends of the energy range and the MECS spectra of many sources as 
bright as X~1822--371 (or even brighter) have been published without any need 
of features at 2 keV or 8 keV.  However, we wanted to be sure that such 
features could not be due to instrumental systematics.  Using a reduced
energy range for the MECS (3--8 keV) these features are still statistically 
significant, giving probabilities of chance improvement of the fit of 
$\sim 5 \times 10^{-3}$ and $\sim 1.8 \times 10^{-5}$ for the low-energy line 
and the absorption edge, respectively.  
We also tried to fit the BeppoSAX spectrum using the whole MECS range 
(1.8--10 keV) and ignoring all the HPGSPC points below
10 keV. Again the addition of the edge at 8.5 keV was statistically 
significant ($\sim 10^{-5}$) demonstrating that this feature is present in 
both the MECS and the HPGSPC data.  We therefore conclude that these features,
which are also expected to be emitted in photo-ionized ADCs
(e.g.\ Ko \& Kallman 1994; Kallman et al.\ 1996), are most probably real 
and not instrumental effects.

The values of the parameters corresponding to
the best fit model are reported in Table 1 (Model 3). We plotted in
Figure 2 (lower panel) the residuals in unit of $\sigma$ with respect
to Model 3, and in Figure 3 the unfolded spectrum corresponding to
this model.

In order to obtain some information about the seed-photon temperature
of the Comptonization spectrum we tried the {\it Comptt} model
(Titarchuk, 1994) instead of the {\it Compst} model. We obtain an
equivalently good fit (see Table~1, Model~4). However, because the
seed-photon temperature is close to the low energy end of our spectral
range, we can only estimate an upper limit to this temperature of
$\sim 0.2$ keV.

To summarize, a brief description of the best fit parameters follows.
We obtained a hydrogen equivalent column $N_{\rm H} \simeq 1.2 \times
10^{21}$ cm$^{-2}$. The hydrogen equivalent column of the partial
covering is $N_{\rm H_{PC}} \simeq 4.5 \times 10^{22}$ cm$^{-2}$ and the
covered region corresponds to a fraction of $71 \%$ of the total.  The
Comptonized component has an electronic temperature of $kT_{\rm e}
\sim 4.7$ keV and optical depth $\tau \sim 14$ for a spherical
geometry.  We find a broad emission line at $\sim 6.5$ keV, with
$FWHM=0.70$ keV and and equivalent width of $\sim 150$ eV. An emission
line at $\sim 1.9$ keV, with $FWHM=0.27$ keV and equivalent width of
$\sim 54$ eV, and an absorption edge at $\sim 8.7$ keV, with optical
depth $\tau_{\rm max} \sim 0.1$, are also detected with high
statistical significance.

\section{Discussion}

We analysed data of the dipping source X~1822--371 from a BeppoSAX 43
ks observation in the energy range 0.1--200 keV.  We obtained the best
fit to these data using a partial covered Comptonization model, plus
two emission lines and an absorption edge.  The results of our
spectral analysis of X~1822--371 are discussed in the following.

The equivalent absorption column $N_H$ we obtained is $\sim 1.2 \times
10^{21}$ cm$^{-2}$.  For a distance to the source of 2.5 kpc (Mason \&
Cordova, 1982) the visual extinction in the direction of X~1822--371
is $A_v=0.87 \pm 0.32$ mag (Hakkila et al. 1997).  Using the observed
correlation between visual extinction and absorption column (Predehl
\& Schmitt 1995) we find $N_H = \left(1.02 \pm 0.02 \right) \times
10^{21}$ cm$^{-2}$, this value is much higher than the value obtained
by Parmar et al. (2000) and by using Model~1.  On the other hand the
expected Galactic absorption is in perfect agreement with the value
that we obtained using Model~3 and Model~4.  In our model the
Comptonization spectrum is partially absorbed for the presence of an
excess of matter close to the source, obscuring part of the emission
region. The absorption column of this partial covering is $\sim 4.5
\times 10^{22}$ cm$^{-2}$, an order of magnitude larger than the
Galactic absorption, and covers a fraction of $\sim 71 \%$ of the
source spectrum.

The continuum is well fitted by a Comptonization spectrum, probably
produced in a hot ($k T_{\rm e} \sim 4.7$ keV) region of moderate
optical depth ($\tau \sim 14$, for a spherical geometry) probably
surrounding the neutron star.  Note that in this spectral
deconvolution there is no need of a soft blackbody. This component is
needed to fit the soft emission of the source when the partial covering
is not used, as in Model~1 of Table~1, i.e. the model adopted by
Parmar et al. (2000).  In this case the measured contribution of the
blackbody to the 1--10 keV flux is greater than 40\%.  In this
scenario, as already noted by Parmar et al. (2000), the blackbody
component is most probably emitted in the inner part of the system
(i.e. from an optically thick boundary layer close to the neutron star
surface or from the neutron star itself).  It is therefore unlikely to
be observed directly in a high inclination ($\sim 85^\circ$) source as
X~1822--371.  One can suppose that part of the blackbody emission can
be scattered into the line of sight by the corona.  However, because
the optical depth of this corona, as deduced by fitting its spectrum
with Comptonization models, is $\tau \ga 10$ (see Table~1 and Parmar
et al. 2000), any blackbody spectrum passing through it will be
(almost) completely reprocessed.  Then the blackbody should contribute
a low percentage of the total source luminosity or should not be
present.  We believe that a more reasonable scenario is the one
proposed by our Model~3 and Model~4, in agreement with the optically
thick scenario used by Heinz \& Nowak (2001) to fit simultaneous RXTE
and ASCA observations.

The Comptonized component probably originates in an ADC that 
could be formed by evaporation of the outer layers of the disk illuminated 
by the emission of the central object (White \& Holt, 1982).  The radius
of the corona can be written as $R_c \simeq \left(M_{NS}/M_{\odot}
\right) T_7^{-1} R_{\odot}$ (White \& Holt, 1982), where $M_{NS}$ is
the mass of the compact object, $M_{\odot}$ and $R_{\odot}$ are mass
and radius of the Sun, and $T_7$ is the ADC temperature in units of
$10^7$ K. Under this hypothesis, using the values reported in Table 1
(Model 3), we find that the radius of the ADC is $R_c \simeq 1.8
\times 10^5$ km.  Similar values for the ADC radius are reported by
White \& Holt (1982, $R_c \simeq 2 \times 10^5$ km) and by Heinz \&
Nowak (2001, $R_c \simeq 2.9 \times 10^5$ km) for X~1822--371.  Using
the relation $\tau = \sigma_T N_e R_c$ we can infer the density of the
ADC, where $\tau$ is the optical depth obtained by the fit, $\sigma_T$
is the Thomson cross-section, $N_e$ is the number of particles per
unit volume and $R_c$ is the ADC radius calculated above.  Note that
we are considering $N_e$ constant along the radius of the corona, that
is a rough approximation. Under this hypothesis we find $N_e \simeq
1.15 \times 10^{15}$ cm$^{-3}$.  This value is in line with previous
simulations of ADC (Vrtilek et al., 1993).  In fact, considering an
inclination angle of $85^{\circ}$ (i.e. an angle of $\sim 5^{\circ}$
from the disk plane), they find a density along the line of sight in
the ADC corona of around $10^{15}$ cm$^{-3}$ (see Fig. 2 in Vrtilek et
al., 1993).  Following Frank et al. (1987) and using the orbital 
parameters reported by Parmar et al. (2000), we estimated that the
accretion disk radius is $R_d \sim 4.3 \times 10^{5}$ km, similar
to the value $R_d \simeq 4 \times 10^{5}$ km obtained by White \& Holt
(1982).  The disk radius is therefore larger than the estimated radius
of the ADC, as expected. In particular the ratio
between the accretion disk radius and the coronal radius is
$R_d \simeq 2.4 R_c$.

Our proposed model is also in agreement with the general behavior of
high inclination dipping sources (see e.g.\ Balucinska-Church et al. 2000;
Smale et al. 2000, and references therein).  The model used to describe
their spectra consists of a point-like blackbody emission and a 
Comptonized component from the ADC.  The spectral evolution during the dips, 
when we observe the source emission through the thickened region of the 
accretion disk rim, can be described in terms of a ``progressive covering'' 
given by an absorber moving progressively across the emission regions. While
the blackbody is rapidly absorbed, suggesting that it is emitted by a
compact region (probably from the NS), the Comptonized component is partially 
absorbed, suggesting that its emission region, i.e.\ the ADC, is extended. 
In particular in the case of X~1624--490, Smale et al. (2000) showed that
the ADC has a larger angular size than the absorber and has a height-to-radius
ratio of $\sim 10\%$, with an estimated coronal radius of $\sim 5 \times 
10^{5}$~km (similar to the value we found above for X~1822--371).
The spectrum of X~1822--371 is very similar to these dip spectra, with the
lack of the blackbody component (that is probably completely absorbed) and
a partially covered Comptonized component. Therefore, in this case of very
high inclination, we probably always observe the source emission though
the thickened accretion disk.
The presence of a thickened outer disk is also suggested by the recent 
{\it XMM-Newton} results on EXO~0748--67 (Cottam et al. 2001).  The presence
of a wealth of emission lines and absorption edges in the range between 
0.4 and 1 keV, showing no eclipses or other modulations related to the 
orbital phase and large widths probably due to velocity broadening,
suggests that these low-energy features are emitted in a flared accretion  
disk extending high above the equatorial plane (Cottam et al. 2001).
 
From the spectral fitting of X~1822--371 we obtained that a fraction of
$71\%$ of the Comptonization spectrum is absorbed by an excess of
matter.  We suppose that this matter is close to the equatorial plane,
at the outer rim of the disk.  Considering the area of the ADC,
$A_{ADC} = 4 \pi R_c^2$, and the area of the covered region as
$A_{cov}=4 \pi R_c h$, where $h$ is the height of the absorbing matter
above the disk, we can write:
\begin{equation}
\frac{A_{cov}}{A_{ADC}} = \frac{h}{R_c} \simeq 0.71.
\end{equation}    
From this equation we obtain that the angle subtended by the absorbing
region is $\theta \sim 16^{\circ}$, adopting the disk radius reported above.
This scenario is in agreement with the
lack of any soft component in our spectral deconvolution.  In fact the
inner region could be either reprocessed by the optically thick ADC
or absorbed by the excess of matter at the outer accretion disk
originating the partial covering.
   
Having the temperature of the seed photons for the Comptonization, we
can derive the radius of the seed-photon emitting region.  Following
in't Zand et al. (1999), this radius can be expressed as $R_W=3 \times
10^{4} D [f_{bol}/\left(1+y\right)]^{1/2}/(kT_0)^2$ km, where D is the
distance of the source in kpc, $f_{bol}$ is the unabsorbed flux in
ergs cm$^{-2}$ s$^{-1}$, $kT_0$ is the seed photon temperature in keV,
and $y=4kT_e \tau^2/m_e c^2$ is the relative energy gain due to the
Comptonization.  We obtained from the fit an upper limit for the
seed-photon temperature (see Table~1, Model~4).  Using this upper
limit in the formula above we obtain a lower limit for the seed-photon
radius of $\sim 24$ km. We can suppose that these photons come from
the inner region of the system, as the neutron star or the boundary
layer between the neutron star and the accretion disk, the inner
radius of the accretion disk or both these regions.

Another component of the model is an emission line at 6.5 keV with an
equivalent width of $\sim 150$ eV.  This is probably due to
fluorescence of moderately ionized iron. A possible origin of the
emission line could be Compton reflection of the photons from the ADC
by the accretion disk (George \& Fabian 1991; Matt, Perola, \& Piro
1991).  However, the large value of the equivalent width seems to be
incompatible with this interpretation. In fact, for inclinations
larger than $80^{\circ}$, the iron line equivalent width should have a
value of $\sim 20$ eV (Brandt \& Matt, 1994).  Moreover the iron line
equivalent width has a maximum value of $\sim 130$ eV for an isotropic
source covering half of the sky ($\Omega/2\pi = 1$) as seen by the
reflector and with an inclination angle of the system $i = 0$.  On the
other hand there are indications suggesting that the emission line in
X~1822--371 originates in the ADC.  In fact, according to Vrtilek et
al. (1993), when the iron emission line originates in ADC, its
equivalent width increases with increasing the inclination angle $i$:
for $i \sim 80^{\circ}$ the equivalent width is $\sim 150$ eV (see
Fig. 6 in Vrtilek et al., 1993). This is in agreement with our results
for X~1822--371,    
for which we obtain an equivalent width of the emission iron line of 
$\sim 150$ eV for $i \sim 85^{\circ}$ (the inclination angle of X~1822--371).
Note that the ADC origin (instead of the disk origin) of the iron line is 
also in agreement with our spectral modelling of the continuum, in which 
the emission from the accretion disk is not directly observed. 

The iron line we observe in X~1822--371 is
quite broad ($\sim 0.7$ FWHM). Since we have excluded a disk origin for
this line we cannot explain its broadening with the standard scenario
adopted for Active Galactic Nuclei (AGN) containing massive black holes,
where the broadening of the iron line is thought to be the result of
general relativistic effects in the innermost regions of an accretion
disk.  In the case of coronal origin of the iron line, its width can be
explained by Compton scattering of the line photons in the optically
thick plasma surrounding the central X-ray source.  This produces a
genuinely broad gaussian distribution of line photons, with $\sigma \ga
E_{\it Fe} (k T_e/m_e c^2)^{1/2}$, where $E_{\it Fe}$ is the centroid
energy of the iron line and $k T_e$ is the electron temperature in the
ADC.  More detailed calculations, in which the dependence on the optical
depth is taken into account, show that this effect can explain the width
of the iron line for temperatures of the emitting region of a few keV
(Kallman \& White 1989, see also Brandt \& Matt 1994).  The presence of
several unresolved components from many iron ionization stages (line
blending) can also contribute to the line broadening.  In this case the
single components could be resolved by the new high resolution instruments
on board of Chandra and {\it XMM-Newton}.  The observation of a broad iron
line whose width cannot be explained by relativistic Doppler 
effects in the innermost region of an accretion disc is interesting and
suggests alternative explanations, the most probable of which being
Comptonization, for the line broadening in LMXBs. However, it is important
to observe that Comptonization fails to explain the shape of the line in
AGNs, for which the most probable broadening mechanism is relativistic 
Doppler effects (e.g.\ Ruszkowski et al. 2000; Misra 2001).

We observe another emission line at $\sim 1.9$ keV with an equivalent
width of $\sim 54$ eV. This line could be due to 	 
emission from the L-shell of ionized iron (Fe XXIII-XXIV for plasma densities
of $\sim 10^{11}$ cm$^{-3}$, see Kallman et al., 1996) or from the K-shell
of highly ionized Si or Mg.  The emission
region of this line could be the outer region of the ADC at high
latitude ($ > 15^{\circ}$) where the coronal density is expected to be
around $10^{11}-10^{12}$ cm$^{-3}$ (Vrtilek et al., 1993).  The last
component is an absorption edge at $\sim 8.7$ keV with an optical
depth of $\sim 0.1$.  Following Turner et al. (1992) for a
correspondence between iron edge energy and ionization level, this
edge corresponds to Fe XXIV. This confirms the idea of highly ionized
material present around the compact object.  The best fit value for
the optical depth $\tau_{edge}$, considering the photoionization cross
section for the K--shell of Fe XXIV (Krolik \& Kallman, 1987),
corresponds to a hydrogen column density of $\sim 1.3 \times 10^{23}$
cm$^{-2}$, assuming cosmic abundance of iron. This is two order of
magnitude higher than the measured Galactic absorption and one order
of magnitude higher than the neutral matter responsible of the partial
covering (see Tab. 1, Model~3 and Model~4). From the estimation of the
coronal density $N_e$ reported above, we can derive the corresponding
hydrogen column density in the ADC, that is roughly $2 \times 10^{25}$
cm$^{-2}$.  This suggests that a part of the ADC could be (photo)
ionized and responsible for the presence of both the iron edge and the
low energy emission line.

\section{Conclusions} 

We analysed data from a BeppoSAX observation of X~1822--371 performed
in 1997 September 9 and 10.  The energy spectrum is well described by
a Comptonized spectrum with partial covering, two iron emission lines
and an absorption edge.  The Comptonized spectrum is probably produced
in the ADC, with electron temperature of $\sim 4.7$ keV and with
moderate optical depth ($\tau \sim 14$ for a spherical geometry). The
partial covering could due to an excess of neutral matter placed close
to the equatorial plane at the outer rim of the accretion disk,
forming a thickened outer disk
subtending an angle of $\sim 16^{\circ}$ as seen from the neutron
star.  The high inclination of the source ($\sim 85^{\circ}$) and the
presence of the cloud of neutral matter above the accretion disk does
not allow to observe the direct emission from the neutron star and the inner
accretion disk.  In the spectrum an iron emission line is present at
$\sim 6.5$ keV with a large equivalent width of 150 eV. We showed that
this line cannot come from the accretion disk but it is probably
produced in the ADC.  Another emission line at $\sim 1.9$ keV, with an
equivalent width of 54 eV, and an absorption edge at $\sim 8.7$ keV
are also detected in the spectrum, which could be produced in an
ionized region in the ADC.

\acknowledgments 

This work was supported by the Italian Space Agency
(ASI), by the Ministero della Ricerca Scientifica e Tecnologica
(MURST).

\newpage 
\section*{TABLE} 
 
\begin{table}[h] 
\begin{center} 
\footnotesize 
\caption{\footnotesize Results of the fit of the X~1822--371 spectrum
in the energy band 0.12--200 keV.  Uncertainties are at the 90\%
confidence level for a single parameter.  The blackbody normalization
(N$_{\rm BB}$) is in units of $L_{37}/D_{10}^2$, where $L_{37}$ is the
luminosity in units of $10^{37}$ ergs/s and $D_{10}$ is the distance
to the source in units of 10 kpc.  $k T_{\rm 0}$ and $k T_{\rm e}$ are
the seed-photon temperature and the electron temperature respectively,
$\tau$ is the optical depth of the scattering cloud. The Comptt and
Compst normalization, N$_{\rm comp}$, are defined as in XSPEC
v.10. $N_{H_{PC}}$ indicates the column absorption of the partial
covering and $f$ the covering fraction.  $f_{bol}$ is the unabsorbed
flux in the 0.1--100 keV range of the Comptonized component in units
of ergs cm$^{-2}$ s$^{-1}$.  $EQW_{\rm Fe}$ indicates the equivalent
width of the line at 6.5 keV, E$_{Fe}$ its centroid and I$_{Fe}$ its
intensity in units of photons cm$^{-2}$ s$^{-1}$; $EQW_{\rm LE}$;
E$_{LE}$ and I$_{LE}$ are the same parameters for the low-energy line
at 1.9 keV.  E$_{\rm edge_l}$ indicates the energy of the absorption
edge at low energy and $\tau_{\rm edge_l}$ its relative optical depth,
E$_{\rm edge_h}$ and $\tau_{\rm edge_h}$ are the same parameters for
the edge at high energy.}
\begin{tabular}{l|c|c|c|c}  
\hline \hline 
 Parameter     & Model 1     & Model  2    & Model 3 &Model 4  \\ 
               & BB + Comptt + & PC + Compst + & PC + Compst + & PC + Comptt +\\
            &  Line + Edge  & Line  & Line+Line+Edge& Line+Line+Edge \\ 
\hline 
$N_{\rm H}$ $\rm (\times 10^{21}\;cm^{-2})$   &$0.62 \pm 0.20$ & $1.43 \pm 0.18$
 &$1.23^{+0.16}_{-0.14}$ & $1. 09^{+0.26}_{-0.51}$  \\
 $N_{H_{PC}}$  $\rm (\times 10^{22}\;cm^{-2})$ & --&$4.31 \pm 0.27$ &
              $4.45 \pm 0.26$&      $4.37 \pm 0.31$ \\
$f$    &--& $0.688 \pm 0.013$  & $0.712 \pm 0.016$ & $0.715^{+0.027}_{-0.022}$\\
$k T_{\rm BB}$ (keV)               & $1.590 \pm 0.041$ & -- & -- &--  \\ 
N$_{\rm BB}$             & $0.273 \pm 0.014$ & --& --&--\\ 
$k T_0$ (keV)       & $0.133^{+0.033}_{-0.038}$& --& --&
                 $ < 0.20$ \\ 
$k T_{\rm e}$ (keV)   & $4.526^{+0.067}_{-0.065}$ &$4.802^{+0.069}_{-0.067}$ &
                         $4.724 \pm 0.075 $  & $4.711\pm 0.075$   \\ 
$\tau$                     & $24.22^{+0.87}_{-0.92}$ &$13.52 \pm 0.26$ &
                       $13.81 \pm 0.35$ &  $14.61 \pm 0.37$\\ 
N$_{\rm comp}$ $(\times 10^{-2})$       & $2.09^{+0.21}_{-0.13}$& $7.71 \pm 0.30$
                    &$7.67 \pm 0.36$ & $6.22^{+3.88}_{-1.10}$  \\ 
$f_{bol}$   &$0.98  \times 10^{-9}$ & $1.53 \times 10^{-9}$ & 
            $1.55 \times 10^{-9}$ &   $1.49 \times 10^{-9}$ \\
$E_{\rm Fe}$ (keV)             & $6.528 \pm 0.046$ &$6.523 \pm 0.055$  & 
               $6.521 \pm 0.050$    & $6.520 \pm 0.050$\\  
$\sigma_{\rm Fe}$ (keV)        & $0.292^{+0.065}_{-0.062}$&
         $0.444^{+0.119}_{-0.086}$ & 
                     $0.302^{+0.075}_{-0.068}$ & $0.303^{+0.075}_{-0.068}$ \\
$I_{\rm Fe}$ $(\times 10^{-3})$    & $0.95 \pm 0.14$ &  $1.37^{+0.23}_{-0.19}$&
           $0.97 \pm 0.15$ & $0.97 ^{+0.17}_{-0.14}$ \\ 
EQW$_{\rm Fe}$ (eV)      &$158 \pm 23$ & $222^{+37}_{-31} $  & $153 \pm 25$ &
                                   $152^{+26}_{-22} $ \\
$E_{\rm LE}$ (keV) &--&--& $1.968^{+0.050}_{-0.061}$&
                             $1.961^{+0.052}_{-0.072}$ \\
$\sigma_{\rm LE}$ (keV) &--&--&$0.116^{+0.083}_{-0.087}$ &
                  $0.124^{+0.093}_{-0.087}$ \\
$I_{\rm LE}$ $(\times 10^{-3})$ &--&--& $1.54^{+0.82}_{-0.54}$ &
                  $1.64^{+1.11}_{-0.62}$ \\
EQW$_{\rm LE}$ (eV) &--&--&$53^{+28}_{-18}$ & $57^{+38}_{-21}$\\
E$_{\rm edge_{l}}$ (keV)  &$1.315 \pm 0.056$ & --&--&--\\  
$\tau_{\rm edge_{l}}$ &$0.325^{+0.077}_{-0.070}$&--&--&--\\ 
E$_{\rm edge_{h}}$ (keV)  &-- & -- & $8.69^{+0.24}_{-0.20}$&
 $8.69^{+0.24}_{-0.20}$ \\  
$\tau_{\rm edge_{h}}$ &--&--& $0.097 \pm 0.029$& $0.097 \pm 0.029$\\ 

$\chi^2$/d.o.f.                   & 245/201 & 291/204  & 206/199 & 206/198\\
\hline
\end{tabular}
\end{center}
\end{table}

\clearpage
 
\section*{FIGURES}

\begin{figure}[h!]
\centerline
{\psfig
{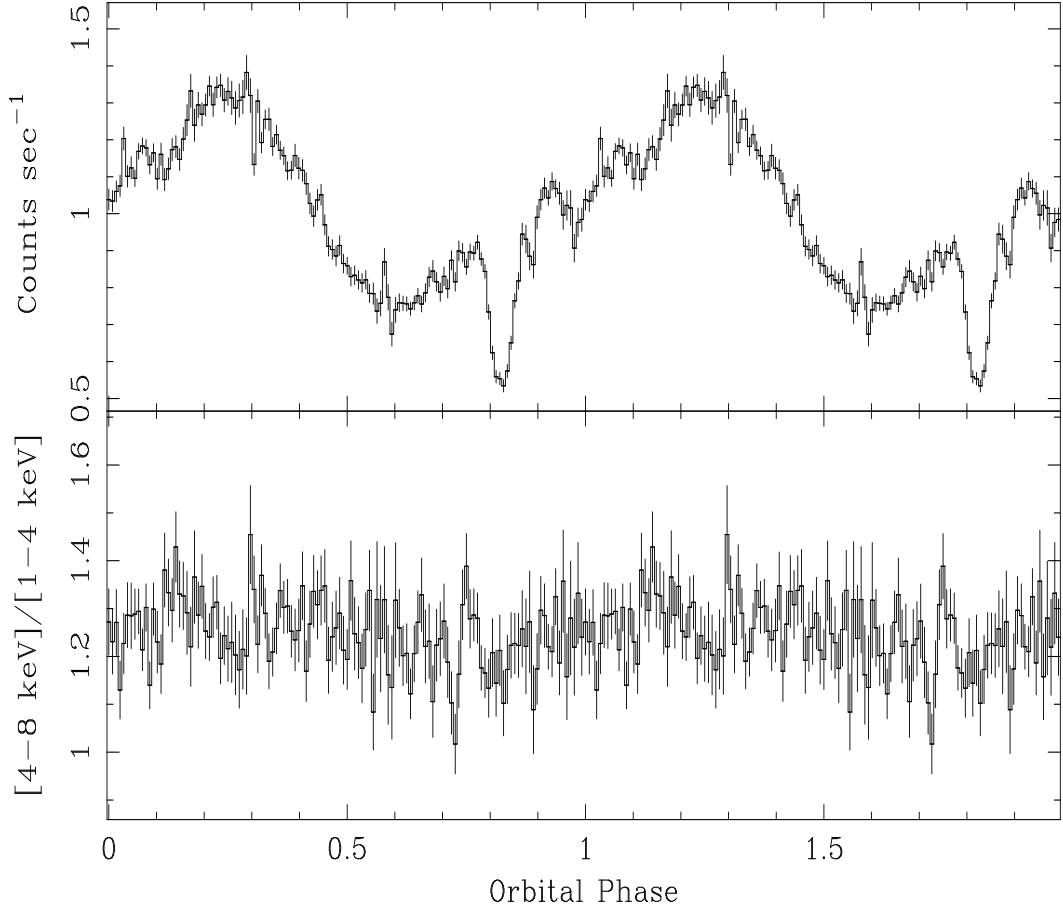}}
\caption{Upper panel: folded light curve of X~1822--371 in the
energy band 1.8--10 keV (MECS data) vs. the orbital phase.  Lower
panel: The ratio of the count rate in the energy band 4--8 keV with
respect to that in 1--4 keV vs. the orbital phase. 
Two orbital phases are shown for clarity.}
\label{fig1}
\end{figure}

\begin{figure}[h!]
\centerline
{\psfig
{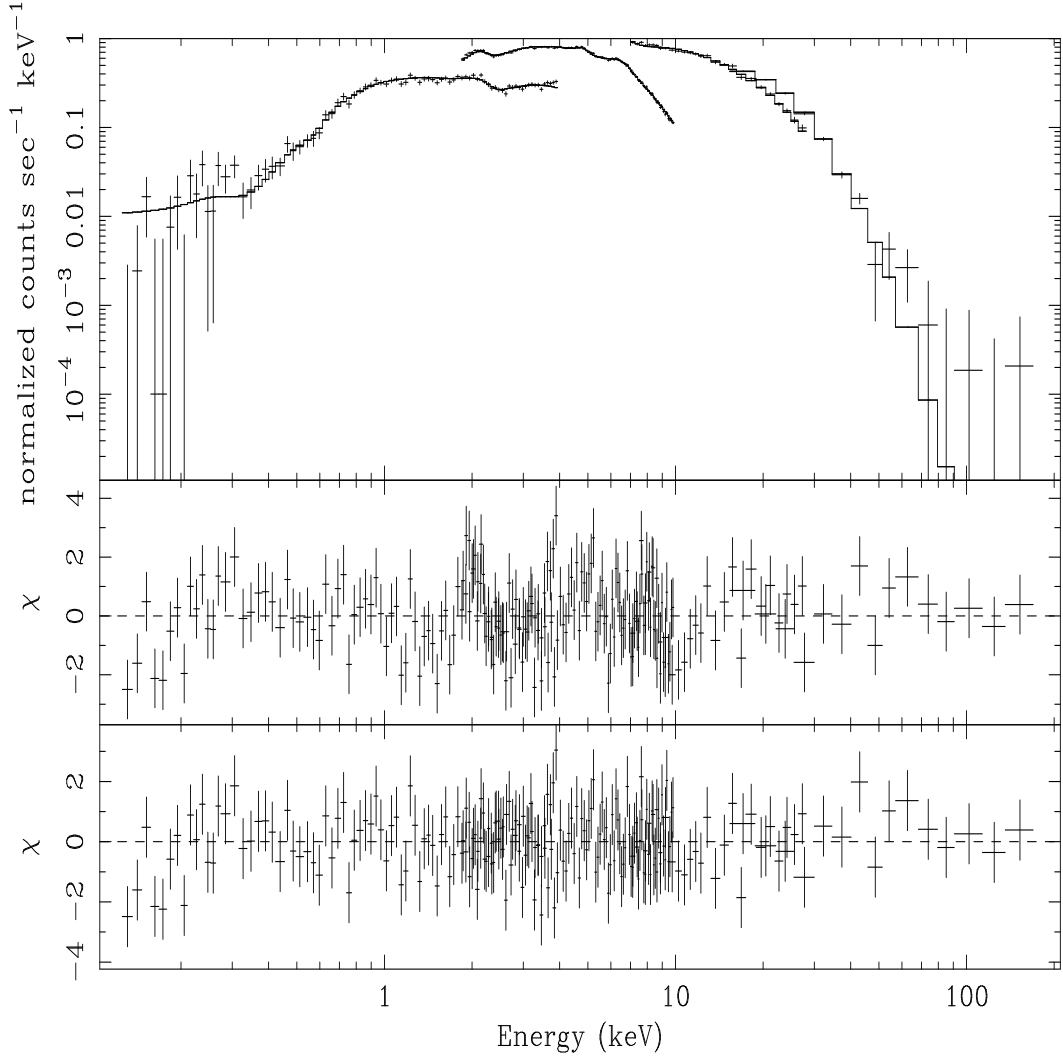}}
\caption{Energy spectra (0.1--200 keV) of X~1822--371.  Data
and the Model~2 (see Tab. 1) are shown in the upper panel, residuals
in units of $\sigma$ with respect to the Model~2 are shown in the
middle panel. In the lower panel the residuals in units of $\sigma$
with respect to the best fit model (Model~3) are shown. }
\label{fig2}
\end{figure}

\begin{figure}[h!]
\centerline
{\psfig
{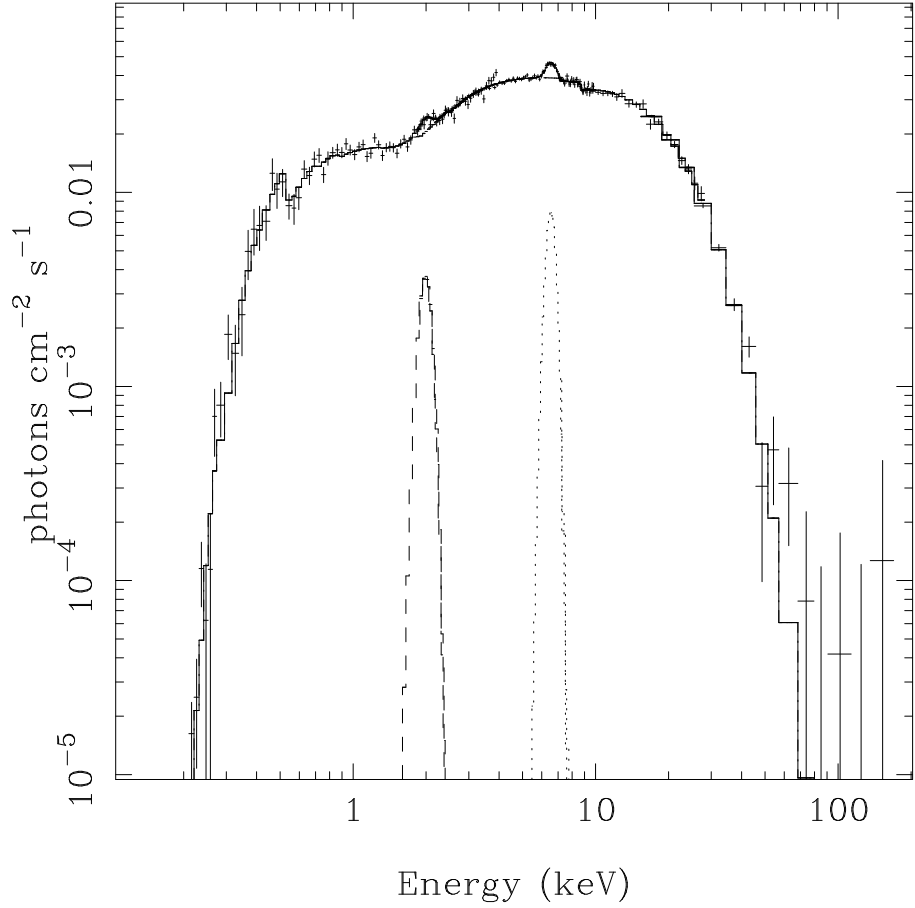}}
\caption{Unfolded spectrum of X~1822--371 and the best fit model 
(Model~3).
The single components of the model are also shown. The solid line is
the {\it Compst} model with partial covering.  The low-energy line at
$\sim 1.9$ keV (dashed line) and the iron emission line at $\sim 6.5$
keV (dotted line) are also shown.  The absorption edge at 8.7 keV is
visible. }
\label{fig3}
\end{figure}

\end{document}